\documentclass[12pt,english]{article}
\usepackage{lmodern}
\usepackage[T1]{fontenc}
\usepackage[latin9]{inputenc}
\usepackage{amsmath}
\usepackage{amssymb}
\usepackage{esint}
\usepackage{geometry}
\makeatletter
\usepackage{float}
\usepackage{hyperref}
\usepackage{cite}
\usepackage{babel}
\usepackage{mathdots}
\usepackage{amsfonts}
\usepackage{mathrsfs}
\usepackage{physics}
\usepackage{amsmath}
\usepackage{esint}
\usepackage{graphicx}
\usepackage{youngtab}
\usepackage{multicol}
\usepackage{multirow}
\usepackage{slashed}
\usepackage{bm}
\usepackage{relsize}
\usepackage[dvipsnames,table,xcdraw]{xcolor}
\usepackage{subfigure}
\usepackage{feynmp}
\usepackage{cancel}
\allowdisplaybreaks
\makeatother
\usepackage{babel}
\hypersetup{
colorlinks=true,
linktoc=page,    
linkcolor=blue,
citecolor=blue,
urlcolor=blue,
colorlinks=true,
}

\begin{document}

\title{Soft theorems in de Sitter spacetime}

\author{Pujian Mao and Kai-Yu Zhang}

\date{}

\def\mytitle{Soft theorems in de Sitter spacetime}

\addtolength{\headsep}{4pt}

\begin{centering}

  \vspace{1cm}

  \textbf{\Large{\mytitle}}

  \vspace{1.5cm}

  {\large Pujian Mao and Kai-Yu Zhang }

\vspace{1cm}

\begin{minipage}{.9\textwidth}\small \it  \begin{center}
     Center for Joint Quantum Studies and Department of Physics,\\
     School of Science, Tianjin University, 135 Yaguan Road, Tianjin 300350, China
 \end{center}
\end{minipage}

\vspace{0.3cm}

\end{centering}

\begin{center}
Emails:\,pjmao@tju.edu.cn,\, kaiyu\_zhang@tju.edu.cn
\end{center}

\vspace{0.2cm}

\begin{center}
\begin{minipage}{.9\textwidth}
\textsc{Abstract}: In this paper, we derive a soft photon theorem and a soft gluon theorem in the de Sitter spacetime from the Ward identity of the near cosmological horizon large gauge transformation. Taking the flat limit of the de Sitter spacetime, the soft theorems naturally recover the corresponding flat spacetime soft theorems.

\end{minipage}

\end{center}

\thispagestyle{empty}

\newpage
\tableofcontents

\section{Introduction}

In the last few years there have been renewed interests on soft theorems. The new enthusiasm comes from a purely theoretical side which resides in the connection between soft theorems
and asymptotic symmetries \cite{Strominger:2013lka,Strominger:2013jfa,He:2014laa,He:2014cra,Strominger:2017zoo}. Soft theorems are nothing but the Ward identities of asymptotic symmetries. Since asymptotic
symmetries do not preserve the vacuum of the theory, they are spontaneously broken. Consequently, soft particles are the Goldstone bosons associated to the (broken) asymptotic symmetries. Because any asymptotically flat spacetime shares the same boundary structure, Hawking, Perry, and Strominger (HPS) argued that black holes will carry soft hairs when acting asymptotic symmetry transformations on stationary black hole solutions \cite{Hawking:2016msc}. The soft hairs are soft particles on the black hole horizon which are just the Goldstone bosons of asymptotic symmetries. The newly discovered black hole soft degree of freedom provides a possible resolution of the black hole information paradox. While, in the original proposal of HPS, the horizon soft particles are created from the point of view of a broken symmetry, the relevance of the soft particles to soft theorems in curved spacetime was examined in \cite{Cheng:2022xyr,Cheng:2022xgm}. The soft theorems in curved spacetime \cite{Cheng:2022xyr,Cheng:2022xgm} are derived from the Ward identity of the near black hole horizon symmetry. In this paper, we will extend the work of \cite{Cheng:2022xyr,Cheng:2022xgm} to the cosmological horizon in de Sitter (dS) spacetime.

There are many reasons to consider soft theorems or soft limits in dS spacetime. For instance, they provide certain relations or constraints on different point correlation functions in dS spacetime, see, e.g., the investigations in \cite{Maldacena:2002vr,Creminelli:2012ed,Hinterbichler:2013dpa,Kundu:2014gxa,Creminelli:2003iq,Cheung:2007sv,Assassi:2012zq,Kundu:2015xta,Shukla:2016bnu}. In this paper, we will compute soft photon and gluon theorems in dS spacetime from the Ward identity of the near cosmological horizon symmetry. Apart from providing a complementary to the previous studies, such as  \cite{Assassi:2012zq,Bittermann:2022nfh,Armstrong:2022vgl,Bhatkar:2022qhz}, there are several reasons to study the soft theorems from the near horizon symmetry point of view. The cosmological horizon, as a null hypersurface, locally is very similar to the black hole horizon. Naively, the derivation should be just repeating the black hole horizon cases, e.g., in \cite{Cheng:2022xyr,Cheng:2022xgm}. However, the presence of a cosmological horizon is a consequence of the fact that the dS cosmological universe is expanding so fast that there are events which will never be seen by an observer inside. In this sense, the cosmological horizon behaves more like a physical boundary.\footnote{See, for instance, \cite{Teitelboim:2001skl,Gomberoff:2003ea} for considering the cosmological horizon as physical boundary to obtain the thermodynamics of a rotating black hole by the Euclidean Schwarzschild-de Sitter solution. } So the near cosmological horizon analysis should be intuitively connected to the asymptotic analysis near null infinity of the asymptotically flat spacetime. More precisely, there is well defined flat limit from the cosmological solution in the coordinates system that we have adopted \cite{Barnich:2012aw}. In the flat limit, the cosmological horizon somehow becomes the null infinity of the asymptotically flat spacetime. Hence, the soft theorems derived in dS spacetime must recover the flat spacetime ones in the flat limit. This is precisely we will verify in this work.\footnote{In the black hole cases, flat spacetime soft theorems can be recovered from the curved space ones. But the conditions to achieve that are somewhat only with mathematical consistency. Physically, those conditions are not reasonable.}

There are two main parts in this paper. In section \ref{Soft photon}, we derive a soft theorem photon theorem from the Ward identity with respect to the near cosmological horizon symmetry in the static patch of dS spacetime in four dimensions. We consider only interaction of massless particles in the present work. To our interest, the in-state $\ket{\mathrm{in}}$ is defined on $\mathcal{H}^{-}$ and the out-state $\ket{\mathrm{out}}$ is defined on $\mathcal{H}^{+}$, see Fig. \ref{f1} for the notations. We introduce a retarded coordinates system to cover near horizon region for $\mathcal{H}^{-}$ part of the horizon. The near horizon symmetry and the solution space of the Maxwell theory coupled with a generic background source is derived under properly selected gauge and boundary conditions which are adapted from the null infinity case in the flat spacetime. For the $\mathcal{H}^{+}$ part, one simply needs to adopt an advanced coordinate which can be considered as a time-reverse transformation from the retarded coordinate. Since the computations on the $\mathcal{H}^{+}$ part is exactly the same, we will not present the full details. By introduing an antipodal matching on the bifurcation sphere $B$, one can obtain the full near horizon symmetry and solution space on the horizon. Then we derive a soft theorem from the Ward identity,
\begin{align}
    \mel**{\mathrm{out}}{Q_{A}^{+}\mathcal{S}-\mathcal{S}Q_{A}^{-}}{\mathrm{in}}= 0,
\end{align}
where $Q^{+}_{A}$ and $Q^{-}_{A}$ are charges generating the transformation of near horizon symmetry defined on $\mathcal{H}^{+}$ and on $\mathcal{H}^{-}$ respectively. In section \ref{Soft gluon}, we derive a soft gluon theorem for a scalar matter field coupled Yang-Mills theory, again from the Ward identity of the near horizon symmetry. For the soft photon and gluon theorems in dS spacetime, the dS radius has the same effect in the soft factor. Both soft theorems have a clear flat limit, in which case, the flat spacetime soft theorems are naturally recovered. In the last section, we give some concluding remarks.


\section{Soft photon theorem in de Sitter spacetime}
\label{Soft photon}

In this section, we study the near horizon behavior of a Maxwell theory coupled with a background source in four dimensional dS spacetime, which is described by the action
\begin{align}
    S=\int \mathrm{d}^4 x\,\sqrt{-g}\left(\frac{1}{4}F^{\mu\nu}F_{\mu\nu}+J_\mu A^\mu\right)\, ,
\end{align}
and derive the soft theorem both in coordinates space and momentum space.

\begin{figure}[http]
    \centering
\includegraphics[width=0.4\linewidth]{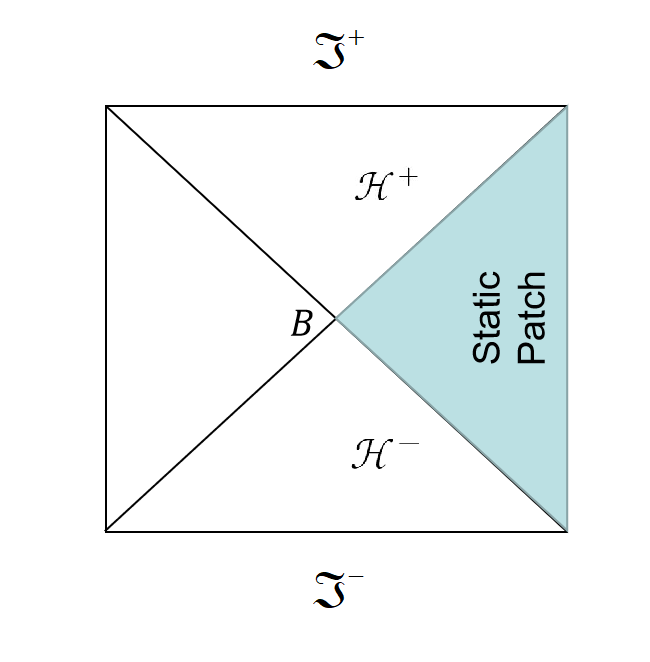}
\caption{Penrose diagram of de Sitter spacetime}
\label{f1}
\end{figure}
We focus on the static patch in dS spacetime, see Fig. \ref{f1} for the Penrose diagram. The line element in static coordinates $(t,r,z,\bar{z})$ is given by
\begin{align}
    &\mathrm{d}s^2=-f(r)\mathrm{d}t^2+f(r)^{-1}\mathrm{d}r^2+\Omega(r)^2\gamma_{z\bar{z}}\mathrm{d}z\mathrm{d}\bar{z}\, ,\\
    &f(r)=\frac{r(2\ell-r)}{\ell^2}\, , \qquad \Omega=\ell-r\, , \qquad \gamma_{z\bar{z}}=\frac{2}{(1+z\bar{z})^2}\, ,
\end{align}
where $\ell$ is the dS radius which is related to the positive cosmological constant by $\Lambda=3\ell^{-2}$ and we have shifted the cosmological horizon at $r=0$. To study the theory in the near horizon region, we introduce the retarded time coordinate $u=t-\frac{1}{2} \ell \log\left(\frac{r}{2\ell -r}\right)$. In the $(u,r,z,\bar{z})$ coordinates, the line element becomes
\begin{equation}\label{metric}
 \mathrm{d}s^2=-f(r)\mathrm{d}u^2-2\mathrm{d}r\mathrm{d}u+\Omega(r)^2\gamma_{z\bar{z}}\mathrm{d}z\mathrm{d}\bar{z}\, ,
\end{equation}
which covers the $\mathcal{H}^{-}$ part of the horizon.

A similar analysis for $\mathcal{H}^{+}$ could be generated by the time-reverse transformation of de Sitter spacetime near the bifurcation sphere $B$, and thus in the rest of this paper, we just show the detail on $\mathcal{H}^{-}$ case. In the retarded coordinates all non-zero components of Christoffel are
\begin{align}
\Gamma _{uu}^{u} & =-\Gamma _{ur}^{r} =-\frac{\Omega }{\ell ^{2}} =-\frac{1}{2} \partial _{r} f & \Gamma _{z\bar{z}}^{u} & =-\gamma _{z\bar{z}} \Omega =\gamma _{z\bar{z}} \Omega \partial _{r} \Omega \\
\Gamma _{uu}^{r} & =\frac{f\Omega }{\ell } =\frac{1}{2} f\partial _{r} f & \Gamma _{z\bar{z}}^{r} & =\gamma _{z\bar{z}} f\Omega =-\gamma _{z\bar{z}} \Omega \partial _{r} \Omega \\
\Gamma _{rz}^{z} & =\Gamma _{r\bar{z}}^{\bar{z}} =-\Omega ^{-1} =\Omega ^{-1} \partial _{r} \Omega  & \Gamma _{zz}^{z} & =\left( \Gamma _{\bar{z}\bar{z}}^{\bar{z}}\right)^{* } =\gamma _{z\bar{z}}^{-1} \partial _{z} \gamma _{z\bar{z}} \, .
\end{align}

\subsection{Gauge fixing, boundary condition and matching}

In this section, we take the radial gauge and the near horizon fall-off conditions for the gauge fields and the conserved current are
\begin{equation}
\begin{aligned}
A_{r} & =0 & A_{u} & =\mathcal{O}( r) & A_{z} & =\mathcal{O}( 1)\\
J_{r} & =0 & J_{u} & =\mathcal{O} ( 1) & J_{z} & =\mathcal{O}( 1).
\end{aligned}
\end{equation}
It is easy to see that the only residual gauge transformation is $\delta A_{\mu}=\partial_\mu \epsilon(z,\bar{z})$, where $\epsilon(z,\bar{z})$ is an arbitrary differentiable function defined on the horizon. Because of the non-vanishing volume in phase space for large gauge symmetry\cite{Miller:2021hty}, this symmetry is non-trivial and leads to the infrared relation between near horizon physics and soft theorem.

The surface charge of the asymptotic symmetry could be constructed by the lower form through the generalized Noether's second theorem \cite{Barnich:2001jy}. We define the surface charge on the bifurcation sphere of the horizon
\begin{equation}
Q_{\epsilon}=\int_B \epsilon\star F ,
\end{equation}
which generates the large gauge transformation in the solution space. In the null infinity case \cite{He:2014cra}, the antipodal matching condition is imposed to preserve the conservation law between two null infinity, because the spatial infinity $i^0$ is not a common boundary of $\mathcal{I}^+$ and $\mathcal{I}^-$. But in this work, horizons live in a finite region, and the bifurcation sphere $B$ is well defined in spacetime without any singularity nor ambiguity. One can consistently study the symmetries near the bifurcation sphere \cite{Adami:2020amw}. The conservation law of the gauge theories should hold automatically without any additional matching condition when $\mathcal{H}^-$ deforms to $\mathcal{H}^+$ as a Cauchy slice. However, we work in the retarded coordinates $(u,r,z,\bar{z})$ or the advanced coordinates $(v,r,z,\bar{z})$ on $\mathcal{H}^+$. Neither of them can really cover the bifurcation point $B$.
Thus, the charges computed in the two coordinates are not connected to each other automatically. To reveal the conservation law on the horizon, we introduce a horizon antipodal matching
\begin{equation}
F_{ru}^{0}(+\infty,0,z,\bar{z})=F_{rv}^{0}(-\infty,0,z,\bar{z})\, ,
\end{equation}
following the standard treatment at null infinity. As one can see from the solution space in the next section, the initial data of the gauge field is completely free. Thus, the matching condition can always be consistently imposed. Then the charge can be defined on the bifurcation sphere $B$ as a limiting case
\begin{equation}
    \begin{aligned}
Q_{\epsilon}=&\lim_{u\rightarrow+\infty}\int_B \mathrm{d}^2z\ell^2\gamma_{z\bar{z}} \epsilon(z,\bar{z})F_{ru}(u,0,z,\bar{z})\\
=&\lim_{v\rightarrow-\infty}\int_B \mathrm{d}^2z\ell^2\gamma_{z\bar{z}} \epsilon(z,\bar{z})F_{rv}(v,0,z,\bar{z})\, .\label{antipodal}
\end{aligned}
\end{equation}

\subsection{Solution space}\label{Soft photon solution}
Under the previously introduced gauge and near horizon fall-off conditions, we can arrange all equations of motion and constraints into several parts.
\begin{itemize}
    \item Continuity equation. We treat the matter current with no dynamics, as a consequence this current must be conserved.
    \begin{align}
        \nabla^\mu J_\mu=0\, .\label{CEA}
    \end{align}
    \item Hyper-surface equation
    \begin{align}
        \nabla_\mu F^{\mu u}=0\, .\label{HEA}
    \end{align}
    \item  Standard equations
    \begin{align}
        \nabla_\mu F^{\mu z}=J^z\, , \ \nabla_\mu F^{\mu \bar{z}}=J^{\bar{z}}\, , \label{SEA}
    \end{align}
    \item  Supplementary equation
    \begin{align}
        \nabla_\mu F^{\mu r}=J^r\, .
    \end{align}
\end{itemize}
The previous gauge fixing does not eliminate all the gauge freedom, the remaining gauge symmetry also connects the equations of motion by the Noether identity
\begin{align}
    \nabla_\nu(J^\nu-\nabla_\mu F^{\mu\nu})=0\, ,
\end{align}
which indicates if the equations \eqref{CEA}--\eqref{SEA} are fulfilled, one have
\begin{align}
\partial_r\left[\sqrt{-g}\left(J^{r}-\nabla_\mu F^{\mu r}\right)\right]=0\, .
\end{align}
The $r$-dependence parts of the supplementary equation are satisfied automatically and the only indeterminate equation is at $\mathcal{O}(1)$. All independent equations of motion yield the solutions as
\begin{align}
J_{u} & =\Omega^{-2}\int _{0}^{r}\mathrm{d} r^{\prime } \gamma _{z\bar{z}}^{-1}( \partial _{z} J_{\bar{z}} +\partial _{\bar{z}} J_z) +\Omega^{-2} J_{u}^{0}\, ,\\
A_{u} & =\int _{0}^{r}\mathrm{d} r^{\prime } \Omega ^{-2}\int _{0}^{r^{\prime }}\mathrm{d} r^{\prime \prime } \gamma _{z\bar{z}}^{-1} \partial _{r^{\prime }}( \partial _{z} A_{\bar{z}} +\partial _{\bar{z}} A_{z}) +\left(\Omega^{-1} -\ell^{-1} \right) A_{u}^{0}\, ,\\
J_{u}^{0} & =-\gamma _{z\bar{z}}^{-1} \Omega ^{-2} \partial _{u}\left( \partial _{z} A_{\bar{z}}^{0} +\partial _{\bar{z}} A_{z}^{0}\right) -\partial _{u} A_{u}^{0}\, ,\\
J_{z} & =-2\partial _{u} \partial _{r} A_{z} +\partial _{r}( f\partial _{r} A_{z}) +\partial _{r} \partial _{z} A_{u} +\Omega^{-2} \partial _{z}\left[\gamma _{z\bar{z}}^{-1}( \partial _{\bar{z}} A_{z} -\partial _{z} A_{\bar{z}})\right]\, ,\\
J_{\bar{z}} & =-2\partial _{u} \partial _{r} A_{\bar{z}} +\partial _{r}( f\partial _{r} A_{\bar{z}}) +\partial _{r} \partial _{\bar{z}} A_{u} +\Omega^{-2} \partial _{\bar{z}}\left[\gamma _{z\bar{z}}^{-1}( \partial _{z} A_{\bar{z}} -\partial _{\bar{z}} A_{z})\right]\, ,
\end{align}
where $A^0_u$ is an integral constant, defined by $A^0_u=\ell^2\partial_r A_u |_{r=0}$, and $A_z^0=A_z |_{r=0}$, $A_{\bar{z}}^0=A_{\bar{z}} |_{r=0}$, $J_u^0=\ell^2J_u |_{r=0}$. One can see from the standard equations that $A_{\bar{z}}^0$ and $A_z^0$ do not have constraints on their time evolution. They are the news functions of this system that indicating the local propagating degree of freedom. It is obvious that $A_u$ is totally determined by the integration constant and $A_z$, $A_{\bar{z}}$ which generate the whole solution space of the near horizon Maxwell theory. Any differentiable scalar function $\epsilon$ defined on a $2$-sphere can be used to transform $A_{\mu}$ through a large gauge transformation of $\delta A_{\mu}=\partial_\mu\epsilon(z,\bar{z})$, resulting in an isomorphism of the solution space.

\subsection{Surface charge of large gauge symmetry}

We express the surface charge as
\begin{align}
    Q^-_\epsilon=\int _{\mathcal{H}^{-}}\mathrm{d}^2 z\mathrm{d} u\ \gamma _{z\bar{z}} \epsilon \left[ \gamma _{z\bar{z}}^{-1} \partial _{u}\left( \partial _{z} A_{\bar{z}}^{0} +\partial _{\bar{z}} A_{z}^{0}\right) +J_{u}^{0}\right]\, ,
\end{align}
where the solutions of the equations of motion are inserted. One can split this charge into two parts
\begin{itemize}
    \item Soft Charge.
    \begin{align}
     Q_S^-=\int _{\mathcal{H}^{-}}\mathrm{d}^2 z\mathrm{d} u\ \epsilon\partial _{u}\left( \partial _{z} A_{\bar{z}}^{0} +\partial _{\bar{z}} A_{z}^{0}\right)\, .
    \end{align}
    If the symmetry parameter $\epsilon$ is a constant at the boundary, the integrand is exactly a total derivative for some functions, and thus this charge is trivial. But for the nontrivial case where $\epsilon$ is a function defined on the horizon, this charge serves a key role in the infrared relation between asymptotic symmetry and the soft theorem.
    \item Hard Charge.
    \begin{align}
     Q^-_H=\int _{\mathcal{H}^{-}}\mathrm{d}^2 z\mathrm{d} u\ \gamma _{z\bar{z}}\epsilon J_{u}^{0}\, .
    \end{align}
    This part acts on $\ket{\mathrm{in}}$ linearly, and the state of incoming particles defined on Hilbert space are eigenstates of the hard charge.
\end{itemize}

The action of the large gauge transformation on the gauge field can be recovered from the canonical commutation relation. The symplectic form of the free Maxwell field, defined on the $\mathcal{H}^{-}$ as a Cauchy surface, is
\begin{equation}
\begin{aligned}
\Omega_{\mathcal{H}^{-}}&=\int_{\mathcal{H}^{-}}\delta A\wedge \delta (\star F)\\
    &=\int_{\mathcal{H}^{-}}\mathrm{d}u\mathrm{d}^2z \delta\partial_uA_{\bar{z}}^{0}\wedge \delta A_{z}^{0}+\delta\partial_uA_z^{0}\wedge\delta A_{\bar{z}}^{0}\, ,
\end{aligned}
\end{equation}
which yields the canonical commutation relation
\begin{align}
    \left[ \partial _{u} A_{\bar{z}}^{0}(u,z,\bar{z}) ,A_{z}^{0}(g,w,\bar{w})\right]=-i\delta(u-g)\delta^2(z-w)\, .
\end{align}
Thus
\begin{align}
     \left[ Q_\epsilon^{-} ,A_{z}^{0}(u,z,\bar{z})\right]=i\partial_{z}\epsilon(z,\bar{z})\, ,
\end{align}
which reveals that $Q_{\epsilon}^{-}$ generates the gauge transformation of near horizon symmetry, as anticipated.

If we assume no magnetic monopole and long-range magnetic force \cite{He:2014cra}, $\mathrm{d}F=0$, $F_{z\bar{z}} |_{u\rightarrow -\infty}=0$, one can find a scalar field $N$ such that
 \begin{align}
        \partial_z N=\int \mathrm{d}u\partial_{u}A_{z}^0\, , \qquad
        \partial_{\bar{z}} N=\int \mathrm{d}u\partial_{u}A_{\bar{z}}^0\,,
    \end{align}
and thus a useful form of soft charge is
  \begin{align}
    Q_S^-=-2\int\mathrm{d}^2z\partial_{\bar{z}}\epsilon\partial_zN\, .\label{soft charge Maxwell}
    \end{align}

\subsection{Mode expansion near the horizon}

Now we derive the mode expansions of the gauge field which are the main ingredients for deriving a soft theorem. It is convenient to write down the mode expansion of the free field operator under the isotropic coordinates $(t,x_1,x_2,x_3)$, in which the line element of the dS spacetime is
\begin{align}
    &\mathrm{d} s^{2}=-\left(\frac{\ell ^{2} -\rho ^{2}}{\ell ^{2} +\rho ^{2}}\right)^{2}\mathrm{d} t^{2} +\mathrm{\left(\frac{2\ell ^{2}}{\ell ^{2} +\rho ^{2}}\right)^{2} d}\vec{x} \cdot \mathrm{d}\vec{x}\, ,\\
t =u+\frac{1}{2} \ell \log&\left(\frac{r}{2\ell -r}\right)\,,\quad x_1=\rho \frac{z+\bar{z}}{z\bar{z}+1}\,, \quad x_2=\rho\frac{z-\bar{z}}{i(z\bar{z}+1)}\,,\quad x_3=\rho\frac{1-z\bar{z}}{z\bar{z}+1}\, ,
\end{align}
where
\begin{align}
    r=\ell\frac{(\ell-\rho)^2}{\ell^2+\rho^2}\, .
\end{align}

In the static patch of dS spacetime, $\partial_t$ is a timelike Killing vector which suggests to define a positive energy of a particle as $\omega=-p_0$. The dispersion relation for massless particles in the isotropic coordinates is
\begin{align}
    -\left(\frac{\rho ^{2} +\ell ^{2}}{\rho ^{2} -\ell ^{2}}\right)^{2} \omega ^{2} +\left(\mathrm{\frac{\ell ^{2} +\rho ^{2}}{2\ell ^{2}}}\right)^{2}\vec{p}^{2}=0\, .
\end{align}
One can find a covariant measure of one-particle phase space
\begin{align}
    \int\frac{\mathrm{d}\omega\mathrm{d}^3\vec{p}}{(2\pi)^3}\delta(p^\mu p_\mu)\Theta(\omega)=\left(\frac{\ell ^{2} -\rho ^{2}}{\ell ^{2} +\rho ^{2}}\right)^{2}\int\frac{\mathrm{d}^3\vec{p}}{(2\pi)^3 2\omega}\, .
\end{align}
For the free massless scalar field $\phi(x)$, we can write this operator as
\begin{align}
    \phi(x^\mu)=\left(\frac{\ell ^{2} -\rho ^{2}}{\ell ^{2} +\rho ^{2}}\right)^{2}\int\frac{\mathrm{d}^3\vec{p}}{(2\pi)^3 2\omega}\left[\mathfrak{a}(p)e^{ip\cdot x}+\mathfrak{a}(p)^{\dagger}e^{-ip\cdot x}\right]\, .
\end{align}
The extra factor $\left(\frac{\ell ^{2} -\rho ^{2}}{\ell ^{2} +\rho ^{2}}\right)^{2}$ comes from the gravitational effect through the dispersion relation. Literally, this mode expansion is the flat spacetime one with the correction from the dispersion relation of the null momentum. The full mode expansion of scalar field in dS spacetime can be performed perturbatively, see, e.g., in \cite{Bhatkar:2022qhz}. However, for the gauge theories, they are the exact mode expansions since the dS spacetime is conformally flat and the gauge theories are conformal invariance in four dimensions at the classical level. The mode expansion for the Maxwell field $A_\mu$ with two polarization vectors that are perpendicular to the direction of propagation is
\begin{align}
    A_{\mu}(x^\mu)=\left(\frac{\ell ^{2} -\rho ^{2}}{\ell ^{2} +\rho ^{2}}\right)^{2}\sum_{\alpha=\pm}\int\frac{\mathrm{d}^3\vec{p}}{(2\pi)^3 2\omega}\left[\epsilon^{*\alpha}_{\mu}a_{\alpha}(p)e^{ip\cdot x}+\epsilon^{\alpha}_{\mu}a_{\alpha}^{\dagger}(p)e^{-ip\cdot x}\right]\, .
\end{align}
One can parameterize the photon momentum as
\begin{align}\label{parametrization}
    p_{\mu }=\frac{\omega }{1+\zeta \bar{\zeta }}\frac{2\ell ^{2}}{\rho ^{2}-\ell ^{2}}\left(\frac{\rho ^{2}-\ell ^{2}}{2\ell ^{2}}( 1+\zeta \bar{\zeta }) ,( \zeta +\bar{\zeta }) ,-i( \zeta -\bar{\zeta }) ,( 1-\zeta \bar{\zeta })\right)\, ,
\end{align}
and similarly construct the polarization vectors as
\begin{align}
\epsilon ^{+\mu } & =\frac{1}{\sqrt{2}}\frac{\ell ^{2} +\rho ^{2}}{2\ell ^{2}}\left(\frac{2\ell ^{2}}{\rho ^{2}-\ell ^{2}}\bar{\zeta } ,1,-i,-\bar{\zeta }\right)\, ,\\
\epsilon ^{-\mu } & =\frac{1}{\sqrt{2}}\frac{\ell ^{2} +\rho ^{2}}{2\ell ^{2}}\left(\frac{2\ell ^{2}}{\rho ^{2}-\ell ^{2}} \zeta ,1,i,-\zeta \right)\, ,
\end{align}
which satisfies
\begin{align}    \epsilon^{\alpha\mu}\epsilon^{\beta*}_{\mu}=\delta^{\alpha\beta}\,, \qquad p_\mu \epsilon^{\alpha\mu}=0\,, \qquad \epsilon^{\alpha}_{r}=0\,.
\end{align}
Projecting those two vectors to the sphere, we have
\begin{align}
\epsilon _{\bar{z}}^{+} & =\frac{2\sqrt{2} \ell ^{2} \rho }{( 1+\zeta \bar{\zeta })\left( \ell ^{2} +\rho ^{2}\right)}\, , & \epsilon _{z}^{-} & =\frac{2\sqrt{2} \ell ^{2} \rho }{( 1+\zeta \bar{\zeta })\left( \ell ^{2} +\rho ^{2}\right)}\, .
\end{align}
To work on the energy representation, we should perform an integral to remove the degree of freedom with respect to the space angle
\begin{equation}
    \begin{aligned}
    \int \frac{\mathrm{d}^{3} p}{2\omega }\epsilon^{*\alpha}_{\mu}a_{\alpha}(p)e^{ip\cdot x}=&\left(\frac{2\ell ^{2}}{\rho ^{2} -\ell ^{2}}\right)^{3} \int \pi \omega \mathrm{d} \omega\epsilon^{*\alpha}_{\mu}(\omega\hat{x}) a_{\alpha}(\omega\hat{x})\\
    &\times\int \sin\mathrm{\theta d} \theta \exp\left[ i\omega \left( -t+\frac{2\ell^2}{\rho^2-\ell^2} \rho  \cos \theta \right)\right]\, .
\end{aligned}
\end{equation}
This allows for an exact computation of the integral,
\begin{equation}
    \begin{aligned}
    \int \sin\mathrm{\theta d} \theta \exp&\left[ i\omega \left( -t+\frac{2\ell^2}{\rho^2-\ell^2} \rho  \cos \theta \right)\right]\\
    =&\frac{1}{\rho\omega }\left(\frac{r}{2\ell -r}\right)^{-i\ell \omega /2}\left(\frac{\rho^2-\ell^2}{2\ell^2}\right)\times e^{-i\omega u}\sin\left[\frac{2\ell^2}{\rho^2-\ell^2} \rho \omega \right]\, .
\end{aligned}
\end{equation}
Finally, the near horizon transverse field operator $A^0_{z}$ is expressed in terms of plane modes
\begin{equation}
\begin{aligned}
A_{z}^{0}( x)  =\frac{2\sqrt{2}}{\pi ^{2}( 1+z \bar{z })}&\frac{\ell ^{6}}{\left[\ell^2+\left(R+\ell\right)^2\right]^3} \int \mathrm{d} \omega  \left(\frac{r}{2\ell -r}\right)^{-i\ell \omega /2}\\
&\times\sin\left[\frac{2\ell^2(R+\ell)\omega}{R(R+2\ell)}\right] \left[e^{\mathrm{-i\omega } u}a_{+}(\omega\hat{x}) +e^{\mathrm{i\omega } u}a_{-}^{\dagger }(\omega\hat{x}) \right]\, ,
\end{aligned}
\end{equation}
where $R=\rho-\ell$, and $R=0$ at the horizon. Following the treatment in \cite{ Cheng:2022xyr,Cheng:2022xgm}, we introduce a near horizon regularization $R\rightarrow R+i\mathcal{R}$ to deal with the divergence from the near horizon limit.

\subsection{Soft photon theorem in coordinates space}

Following previous treatment \cite{He:2014cra,Cheng:2022xyr}, we choose $\epsilon=\frac{1}{z-w}$, which leads to
\begin{align}
    \partial_{\bar{z}}\epsilon(z,\bar{z})=2\pi\delta^2(z-w)\, .
\end{align}
Hence, the soft charge  \eqref{soft charge Maxwell} has the form in terms of annihilation and creation operators
\begin{equation}
    \begin{aligned}
    Q_{S}^{-}
    &=-8\pi^2 i \lim_{\omega\rightarrow0}\left[\omega\widetilde{A_z^0}\right]\, ,
\end{aligned}
\end{equation}
where $\widetilde{A_z^0}$ is defined by Fourier relation $A_z^0=\int_{-\infty}^{+\infty} \mathrm{d}\omega e^{iu\omega}\widetilde{A_z^0}$, and thus
\begin{align}
    Q_{S}^{-}=-\frac{i2\sqrt{2}}{1+w\bar{w}}\frac{\ell^2}{R}\lim_{\omega\rightarrow 0}\left(\omega^2a_{+}(\omega\hat{x})+\omega^2a_{-}^{\dagger}(\omega\hat{x})\right)\, .
\end{align}
The soft charge acts on the in-state as
\begin{equation}
  Q_{S}^{-}\ket{\mathrm{in}}=-\frac{i2\sqrt{2}}{1+w\bar{w}}\frac{\ell^2}{R}\lim_{\omega\rightarrow 0}\omega^2 a_{-}^{\dagger}(\omega\hat{x})\ket{\mathrm{in}}\, .
\end{equation}
Regarding to the action of the hard charge on the state, for simplicity, we consider a complex scalar field source. The conserved current in terms of complex scalar fields is
 \begin{equation}
J^0_{u}=iQ\left(\phi^0\partial_u\bar{\phi}^0-\bar{\phi}^0\partial_u\phi^0\right)\, .
 \end{equation}
The symplectic form of a free scalar theory on the horizon is
\begin{align}
\Omega_{\mathcal{H}^{-}}=\int_{\mathcal{H}^{-}}\mathrm{d}u\mathrm{d}^2z\gamma_{z\bar{z}}\ell^2\left(\delta\partial_{u}\bar{\phi}^{0}_{k}\wedge\delta\phi^{0}_{k}+\delta\partial_u\phi^{0}\wedge\delta\bar{\phi}^{0}_{k}\right)\, ,
\end{align}
which yields
\begin{align}
     \left[ \partial_u \bar{\phi}^{0}_k(u,z,\bar{z}),\phi^{0}_k(g,w,\bar{w})\right]=-i\gamma^{-1}_{z\bar{z}}\ell^{-2}\delta ^{2}( z-w) \delta ( u-g) \, .
\end{align}
Thus the commutation relation between the hard charge and the scalar fields defined on $\mathcal{H}^{-}$ is
\begin{align}
\left[Q_{H}^{-},\phi^{0}_k(u_{k},z_{k},\bar{z}_{k})\right]=Q_k\epsilon(z_{k},\bar{z}_{k})\phi^{0}_{k}(u_{k},z_{k},\bar{z}_{k})\, .\label{Hard Charge Relation Maxell}
\end{align}
Hence
\begin{align}
Q_{H}^{-}\ket{\mathrm{in}}=\sum_{k=1}^{n}Q_k \epsilon(z_k,\bar{z}_{k})\ket{\mathrm{in}}\,.
\end{align}
Then the action of the hard charge on the in-state is
\begin{equation} Q_{H}^{-}\ket{\mathrm{in}}=\sum_{k=1}^{n}\frac{Q_k^{\mathrm{in}}}{z_k^{\mathrm{in}}-w}\ket{\mathrm{in}}\, ,
\end{equation}
Finally, combining those terms and considering also the out-state part, the Ward identity in the coordinates space yields the soft theorem as
\begin{align}
     \lim_{\omega\rightarrow 0} \mel**{\mathrm{out}}{a_{+}\mathcal{S}-\mathcal{S}a_{-}^{\dagger}}{\mathrm{in}}&=i\frac{R}{2\ell^2\omega}\frac{1+w\bar{w}}{\sqrt{2}\omega}\left[\sum_{l=1}^{m}\frac{Q_l^{\mathrm{out}}}{w-z_l^{\mathrm{out}}}-\sum_{k=1}^{n}\frac{Q_k^\mathrm{in}}{w-z_k^{\mathrm{in}}}\right]\mel**{\mathrm{out}}{\mathcal{S}}{\mathrm{in}}\, .
\end{align}

\subsection{Soft theorem in momentum space}

From the null parameterization of the incoming hard momenta in the near horizon region,
\begin{align}
    p_{k\mu }^{\mathrm{in}}=\frac{E_k^{\mathrm{in}}}{1+z_k^\mathrm{in}\bar{z }_k^\mathrm{in}}\frac{2\ell ^{2}}{\rho ^{2}-\ell ^{2}}\left(\frac{\ell ^{2} -\rho ^{2}}{2\ell ^{2}}( 1+z_k^\mathrm{in} \bar{z }_k^\mathrm{in}) ,( z_k^\mathrm{in} +\bar{z }_k^\mathrm{in}) ,-i( z_k^\mathrm{in} -\bar{z }_k^\mathrm{in}) ,( 1-z_k^\mathrm{in} \bar{z }_k^\mathrm{in})\right)\, .
\end{align}
it is easy to show that
\begin{align}
\sum_{k=1}^{n}Q_k\frac{p_k^{\mathrm{in}}\cdot\epsilon}{p_k^{\mathrm{in}}\cdot p^{\mathrm{in}}}=\sum_{k=1}^{n}\frac{R}{\ell }\frac{( 1+w\bar{w})}{\sqrt{2} \omega }\frac{Q_{k}^{\mathrm{in}}}{w-z_{k}^{\mathrm{in}}}\, .
\end{align}
Similar relation can be derived for the outgoing hard momenta. Those relations allow us to write the soft photon theorem in momentum space as
\begin{align}\label{softphoton}
     \lim_{\omega\rightarrow 0} \mel**{\mathrm{out}}{a_{+}\mathcal{S}-\mathcal{S}a_{-}^{\dagger}}{\mathrm{in}}&=\frac{i}{2\omega\ell}\left[\sum_{l=1}^{m}Q_l^{\mathrm{out}}\frac{p_l^{\mathrm{out}}\cdot\epsilon}{p_l^{\mathrm{out}}\cdot p}-\sum_{k=1}^{n}Q_{k}^{\mathrm{in}}\frac{p_k^{\mathrm{in}}\cdot\epsilon}{p_k^{\mathrm{in}}\cdot p}\right]\mel**{\mathrm{out}}{\mathcal{S}}{\mathrm{in}}\, .
\end{align}

\subsection{Flat limit}

A limit of the spacetime geometries is in general a subtle issue \cite{Geroch:1969ca}. For instance, the flat limit $\ell\to\infty$ is not well defined for the line element \eqref{metric}. While, the flat limit is well defined for the case using $\Omega$ as the radial coordinate, where the line element is
\begin{equation}
\mathrm{d}s^2=(\frac{\Omega^2}{\ell^2}-1)\mathrm{d}u^2-2\mathrm{d}\Omega\mathrm{d}u+\Omega^2\gamma_{z\bar{z}}\mathrm{d}z\mathrm{d}\bar{z}\, , \quad \Omega=\ell - r \, .
\end{equation}
The location of the cosmological horizon in term of $\Omega$ is simply $\Omega=\ell$. Hence, the flat limit yields that the cosmological horizon becomes the null infinity of the flat spacetime.

Regarding to the flat limit of the soft theorem in dS spacetime, it is more reasonable to consider the momentum space one due to the subtleties from the coordinates in the spacetime limit. It is obvious from \eqref{softphoton} that the soft theorem in the flat limit recovers the flat spacetime soft photon theorem. The only condition is that the dS radius multiplied by the soft energy is a finite constant in the flat limit, see also \cite{Hijano:2020szl} for relevant discussions about the flat limit of AdS spacetime and soft limit. This constraint can be understood as follows. In terms of the cosmological constant $\Lambda$, the flat limit is only physically reasonable when $\Lambda$ approaching zero as fast as the soft energy $\omega$. Note that any constant prefactor can be absorbed in the null parametrization \eqref{parametrization} and the imaginary unit $i$ would not change the physical observable cross section. The imaginary unit $i$ can also be understood as a regularization in the flat limit in terms of $\Lambda$. Similar to the black hole cases \cite{Cheng:2022xyr,Cheng:2022xgm}, $\Lambda=0$ means there is no horizon in the spacetime at all. Since our derivation is based on the near horizon symmetry, the flat limit should not be taken directly, but by introducing a regularization $\Lambda\to\Lambda + i \lambda$. Then the condition for $\lambda$ in the flat limit is that it should be proportional to the soft energy $\omega$.


\section{Soft gluon theorem in de Sitter spacetime}
\label{Soft gluon}

In this section, we explore the soft gluon theorem from the Ward identity with respect to the near horizon symmetry of a scalar matter field coupled Yang-Mills theory in dS spacetime. For generality, we do not specify the
gauge group. The theory is described by the action
\begin{align}
    S=\int \mathrm{d}^4 x\,\sqrt{-g}\left(\frac{1}{4}F^{a\mu\nu}F^a_{\mu\nu}+D_\mu\phi^a D^\mu\bar{\phi}^a\right)\, .
\end{align}
The scalar fields $\phi^a$ are in the adjoint representation of the gauge group,
\begin{align}
&\left[ T^{a} ,T^{b}\right]  =C^{abc} T^{c}\, ,\qquad  \left( T^{a}\right)_{bc}  =C^{abc}\, , \qquad \ ( D_{\mu } \phi)^{a}=\nabla _{\mu } \phi^{a} +ig_{YM} Q C^{abc} A_{\mu }^{b} \phi^{c}\,
\end{align}
the gauge field strength is defined by
\begin{align}
    &F^a_{\mu\nu}=\partial _{\mu } A_{\nu }^{a} -\partial _{\nu } A_{\mu }^{a} +ig_{YM} C^{abc} A_{\mu }^{a} A_{\nu }^{b}\,.
\end{align}
The gauge transformation satisfies the symmetry algebra
\begin{align}    \left[\delta_{\epsilon_1},\delta_{\epsilon_2}\right]A^a_\mu=\delta_{\left[{\epsilon_1},{\epsilon_2}\right]}A^a_\mu\, , \qquad \left[\delta_{\epsilon_1},\delta_{\epsilon_2}\right]\phi^a=\delta_{\left[{\epsilon_1},{\epsilon_2}\right]}\phi^a\, .
\end{align}

\subsection{Gauge fixing and boundary Condition}

We choose the radial gauge and the following near horizon fall-off conditions,
\begin{equation}
\begin{aligned}
A_{r}^a  =0 \, ,\quad  A_{u}^a  =\mathcal{O}( r) \, ,\quad   A_{z}^a  =\mathcal{O}( 1)\, ,\quad   \phi^a=\mathcal{O}( 1)\, ,\quad  \bar{\phi}^a=\mathcal{O}( 1)\, .
\end{aligned}
\end{equation}
The only residual gauge transformation is $\delta A_{\mu}^{a}=D_\mu \epsilon^a(z,\bar{z})$ where $\epsilon^a(z,\bar{z})$ are arbitrary differentiable functions defined on $S^2$. The surface charge is constructed by \cite{Mao:2017tey}
    \begin{equation}
    \begin{aligned}
       Q_{\epsilon}=&2\int_B \Tr[\epsilon\star F]\\
        =&\int_B \mathrm{d}^2z \Omega^2\gamma_{z\bar{z}}\epsilon^aF^{a}_{ru}\, ,
    \end{aligned}
    \end{equation}
which generates the nontrivial large gauge transformation in the solution space of the near horizon scalar fields coupled Yang-Mills theory.

\subsection{Solution space}

Following the analysis of the Abelian case, we divide equations of motion into three parts
\begin{itemize}
    \item Hyper-surface equation
    \begin{align}
        D_\mu F^{a\mu u}=0\, .\label{HEN}
    \end{align}
    \item Standard equations
    \begin{align}
        D_\mu F^{a\mu z}=J^{a z}\, , \ D_\mu F^{a\mu \bar{z}}=J^{a \bar{z}}\, , \ D^\mu D_\mu\phi^a=D^\mu D_\mu\bar{\phi}^a=0\, .\label{SEN}
    \end{align}
    \item  Supplementary equation
    \begin{align}
        D_\mu F^{\mu r}=J^r\, .\label{SUEN}
    \end{align}
\end{itemize}
The conserved current is given by
\begin{align}
    J_{\mu}^a=iQC^{abc}\left( \phi ^{b} D_{\mu }\bar{\phi }^{c} -\bar{\phi }^{b} D_{\mu } \phi ^{c}\right)\, .
\end{align}
The Noether identity
\begin{align}
D_{\mu }\left( g_{YM} J^{a\mu } -D_{\nu } F^{a \nu \mu }\right) & =ig_{YM} QC^{abc}\left[ \phi ^{b} D_{\mu } D^{\mu }\bar{\phi }^{c} -\bar{\phi }^{b} D_{\mu } D^{\mu } \phi ^{c}\right]\, ,
\end{align}
yields that, once equations \eqref{HEN} and \eqref{SEN} are satisfied, we have
\begin{align}
 \partial _{r}\left[\sqrt{-g} \left( D_{\mu }F^{a\mu r} -g_{YM}J^{ar}\right)\right] & =0\, ,
\end{align}
which means only the $\mathcal{O}(1)$ equation is indeterminate. We present all necessary equations
\begin{itemize}
    \item Standard equations
\begin{equation}
\begin{aligned}
\mathcal{D}^{A}\mathcal{D}_{A}\bar{\phi }^{a} +\Omega^{-1}& \left[ \partial _{r}\left( \Omega f\partial _{r}\bar{\phi }^{a}\right) +f\partial _{r} \Omega \partial _{r} \bar{\phi}^{a} \right] -2\Omega^{-1} \partial _{r} \partial _{u}\left( \Omega \bar{\phi }^{a}\right)\\
&-ig_{YM} C^{abc}\left(h^{AB}  A_{A}^{b} D_{B}\bar{\phi }^{c} +fA_{r}^{b} \partial _{r}\bar{\phi }^{c} -A_{u}^{b} \partial _{r}\bar{\phi }^{c}\right)\\
 &\quad \quad \quad +ig_{YM} C^{abc}\left[\mathcal{D}_{A}\left( A^{bA }\bar{\phi }^{c}\right) -\Omega^{-2} \partial _{r}\left( \Omega ^{2} A_{u}^{b}\bar{\phi }^{c}\right)\right] =0\, ,
\end{aligned}
\end{equation}
\begin{equation}
\begin{aligned}
\mathcal{D}^{A}\mathcal{D}_{A}\phi^{a} +\Omega^{-1}& \left[ \partial _{r}\left( \Omega f\partial _{r}\phi^{a}\right) +f\partial _{r} \Omega \partial _{r} \phi^{a} \right] -2\Omega^{-1} \partial _{r} \partial _{u}\left( \Omega \phi^{a}\right)\\
&+ig_{YM} C^{abc}\left(h^{AB}  A_{A}^{b} D_{B}\phi^{c} +fA_{r}^{b} \partial _{r}\phi^{c} -A_{u}^{b} \partial _{r}\phi^{c}\right)\\
 &\quad \quad \quad -ig_{YM} C^{abc}\left[\mathcal{D}_{A}\left( A^{bA }\phi^{c}\right) -\Omega^{-2} \partial _{r}\left( \Omega ^{2} A_{u}^{b}\phi^{c}\right)\right] =0\, ,
\end{aligned}
\end{equation}
\begin{equation}
\begin{aligned}
\partial _{r} \partial _{z} A_{u}^{a}-2\partial _{u} \partial _{r} A_{z}^{a}  +\partial _{r}\left( f\partial _{r} A_{z}^{a}\right)& -\partial _{z}\left( F_{\ \ \ z}^{a\bar{z}}\right)-g_{YM}J_{z}^a \\
= ig_{YM} C^{abc} & \left[ A_{\bar{z}}^{b} \partial _{r} A_{z}^{c} +A_{z}^{b} F_{\ \ \ z}^{c\bar{z}} +A_{z}^{b} \partial _{r} A_{\bar{z}}^{c} -2A_{u}^{b} \partial _{r} A_{z}^{c}\right]\, ,
\end{aligned}
\end{equation}
\begin{equation}
\begin{aligned}
\partial _{r} \partial _{\bar{z}} A_{u}^{a}  -2\partial _{u} \partial _{r} A_{\bar{z}}^{a}  +\partial _{r}\left( f\partial _{r} A_{\bar{z}}^{a}\right)& -\partial _{\bar{z}}\left( F_{\ \ \ \bar{z}}^{az}\right)-g_{YM}J_{\bar{z}}^a\\
= ig_{YM} C^{abc} & \left[ A_{z}^{b} \partial _{r} A_{\bar{z}}^{c} +A_{\bar{z}}^{b} F_{\ \ \ \bar{z}}^{cz} +A_{\bar{z}}^{b} \partial _{r} A_{z}^{c} -2A_{u}^{b} \partial _{r} A_{\bar{z}}^{c}\right]\, ,
\end{aligned}
\end{equation}
where $\mathcal{D}^A$ and $h_{AB}$ are the covariant derivative and metric defined on the horizon respectively. It is clear that  $A^{a0}_z=A^a_z |_{r=0} $, $ A^{a0}_{\bar{z}}=A^a_{\bar{z}} |_{r=0}$, $ \phi^{a0}=\phi^a|_{r=0}$ do not have constraints for their time evolution. They are the news functions of the system that characterize the local propagating degree of freedom.

\item Hyper-surface equation
\begin{equation}
    \begin{aligned}
A_{u}^{a} & =\int _{0}^{r}\mathrm{d} r^{\prime } \Omega ^{-2}\int _{0}^{r^{\prime }}\mathrm{d} r^{\prime \prime } \gamma _{z\bar{z}}^{-1} \partial _{r^{\prime \prime }}\left( \partial _{\bar{z}} A_{z}^{a} +\partial _{z} A_{\bar{z}}^{a}\right) +\left(\ell^{-1} -\Omega^{-1}\right) A_{u}^{a0}\\
 & \ \ \ \ \ \ \ \ \ \ \ \ \ \ \ \ \ \ +\int _{0}^{r}\mathrm{d} r^{\prime } \Omega ^{-2}\int _{0}^{r^{\prime }}\mathrm{d} r^{\prime \prime } \gamma _{z\bar{z}}^{-1} ig_{YM} C^{abc}\left( A_{z}^{b} \partial _{r^{\prime \prime }} A_{\bar{z}}^{c} +A_{\bar{z}}^{b} \partial _{r^{\prime \prime }} A_{z}^{c}\right)\, ,
\end{aligned}
\end{equation}
where $A^{a0}_u$ is an integral constant and defined by $A^{a0}_u=\ell^2\partial_rA^a_u |_{r=0}$.

\item Supplementary equation
\begin{equation}
    \begin{aligned}
\partial _{u} A_{u}^{a0} =-\gamma _{z\bar{z}}^{-1} \partial _{u} & \left( \partial _{\bar{z}} A_{z}^{a0} +\partial _{z} A_{\bar{z}}^{a0}\right)\\
&-ig_{YM}  C^{abc} \gamma _{z\bar{z}}^{-1}\left( A_{z}^{0b} \partial _{u} A_{\bar{z}}^{0c} +A_{\bar{z}}^{0b} \partial _{u} A_{z}^{0c}\right)\\
&\quad\quad\quad\quad\quad\quad - iQg_{YM} C^{abc} \Omega ^{2}\left( \phi ^{0b} \partial _{u}\bar{\phi }^{0c} -\bar{\phi }^{0b} \partial _{u} \phi ^{0c}\right)\, .
\end{aligned}
\end{equation}
\end{itemize}
The full solution space of the non-Abelian theory is generated by the integration constant $A^{a0}_u$ and the initial data $\{A_{z}^a, A_{\bar{z}}^a,\phi^a,\bar{\phi}^a\}$ and the news functions at any time.

\subsection{Surface charge of large gauge symmetry}

Inserting the near horizon solution, the boundary charge is reduced to
\begin{equation}
    \begin{aligned}
Q_\epsilon=&\int_{\mathcal{H}^{-}}\mathrm{d}^2z\mathrm{d}u\epsilon^a\partial_u\left(\partial_{\bar{z}}A_z^{a0}+\partial_zA_{\bar{z}}^{a0}\right)\\
&+ig_{YM} C^{abc}\int_{\mathcal{H}^{-}}\mathrm{d}^2z\mathrm{d}u\epsilon^a\left[\left( A_{z}^{a0} \partial _{u} A_{\bar{z}}^{0b} +A_{\bar{z}}^{a0} \partial _{u} A_{z}^{0b}\right)\right.\\
&\quad \quad \quad \quad \quad \quad \quad \quad \quad \quad  \left.+ Q \gamma _{z\bar{z}} \Omega ^{2}\left( \phi ^{0b} \partial _{u}\bar{\phi }^{0c} -\bar{\phi }^{0b} \partial _{u} \phi ^{0c}\right)\right]\, .
    \end{aligned}
\end{equation}
We decompose this charge into two parts
\begin{itemize}
  \item Soft Charge
     \begin{align}
    Q_S^{-}=-2\int_B\mathrm{d}^2z\partial_{\bar{z}}\epsilon^a\partial_zN^a\, ,
    \end{align}
    where
    \begin{align}
        \partial_z N^a=\int\mathrm{d}u\partial_u A_{z}^{a0}\, ,\ \partial_{\bar{z}} N^a=\int\mathrm{d}u\partial_uA_{\bar{z}}^{a0}\,  .
    \end{align}
The soft charge is very similar to the case of the Maxwell theory. The only adaption is the group index on the gauge field.
    \item Hard Charge
    \begin{equation}
    \begin{aligned}
       Q_H^{-}=&ig_{YM} C^{abc}\int_{\mathcal{H}^{-}}\mathrm{d}^2z\mathrm{d}u\epsilon^a\left[\left( A_{z}^{0b} \partial _{u} A_{\bar{z}}^{0c} +A_{\bar{z}}^{b0} \partial _{u} A_{z}^{0c}\right)\right.\\
&\quad \quad \quad \quad \quad \quad \quad \quad \quad \quad  \left.+ Q\gamma _{z\bar{z}} \Omega ^{2}\left( \phi ^{0b} \partial _{u}\bar{\phi }^{0c} -\bar{\phi }^{0b} \partial _{u} \phi ^{0c}\right)\right]
    \end{aligned}
    \end{equation}
The first line is zero in the Abelian case since a linearization will turn off the self-interacting pieces.
\end{itemize}
The canonical commutation relations of the Abelian case on $\mathcal{H}^{-}$ can be generalized to the present case as
\begin{align}\label{Acommutator}
     \left[ \partial _{u} A_{\bar{z}}^{0a} (u,z,\bar{z}),A_{z}^{0b}(g,w,\bar{w})\right]=-i\delta ^{2}( z-w) \delta ( u-g) \delta ^{ab}\, ,
 \end{align}
 thus
 \begin{align}
      \left[ Q^{-}_{\epsilon},A_{z}^{0d}(u,z,\bar{z})\right]=i\left[\partial_\mu\epsilon^d(z,\bar{z})+ig_{YM}C^{abd}A^{0a}_{z}\epsilon^b(z,\bar{z})\right]\, ,
 \end{align}
 which is the infinitesimal gauge transformation of the gauge field.

 \subsection{Mode expansion near the horizon}

The mode expansion of the Yang-Mills field is almost the same as the Maxwell field with the only adaption of a group index. We just give the final results of the mode expansion here without repeating the same type of computation. The mode expansion near the horizon field is given by
 \begin{equation}
\begin{aligned}
A_{z}^{0a}( x)  =\frac{2\sqrt{2}}{\pi ^{2}( 1+z \bar{z })}&\frac{\ell ^{6}}{\left[\ell^2+\left(R+\ell\right)^2\right]^3} \int \mathrm{d} \omega  \left(\frac{r}{2\ell -r}\right)^{-i\ell \omega /2}\\
&\times\sin\left[\frac{2\ell^2(R+\ell)\omega}{R(R+2\ell)}\right] \left[e^{\mathrm{-i\omega } u}a_{+}^{a}(\omega\hat{x}) +e^{\mathrm{i\omega } u}a_{-}^{a\dagger }(\omega\hat{x}) \right]\, .
\end{aligned}
\end{equation}

\subsection{Soft gluon theorem and flat limit}

Following the choice in \cite{Cheng:2022xgm}, we choose $\epsilon^a(z,\bar{z})=\frac{1}{z-w}$ where $a$ denotes a particular component of $\epsilon$ and the rest components are set to zero. Then, the soft charge of the Yang-Mills theory in terms of the annihilation and creation operators defined in the mode expansion is
\begin{align}
    Q_{S}^{-}=-\frac{i2\sqrt{2}}{1+w\bar{w}}\frac{\ell^2}{R}\lim_{\omega\rightarrow 0}\left(\omega^2a_{+}^{a}(\omega\hat{x})+\omega^2a_{-}^{a\dagger}(\omega\hat{x})\right)\, .
\end{align}
Acting the soft charge on in-state yields
\begin{align}
  Q_{S}^{-}\ket{\mathrm{in}}&=-\frac{i2\sqrt{2}}{1+w\bar{w}}\frac{\ell^2}{R}\lim_{\omega\rightarrow 0}\omega^2 a_{-}^{a\dagger}(\omega\hat{x})\ket{\mathrm{in}}\, .
\end{align}
Regarding to the hard part, one needs to compute the commutation relation from the symplectic form of $\phi^a$, which is given by
 \begin{align}
\Omega_{\mathcal{H}^{-}}=\int_{\mathcal{H}^{-}}\mathrm{d}u\mathrm{d}^2z\gamma_{z\bar{z}}\ell^2\left(\delta\partial_{u}\bar{\phi}^{0a}\wedge\delta\phi^{0a}+\delta\partial_u\phi^{0a}\wedge\delta\bar{\phi}^{0a}\right)\, .
 \end{align}
Hence,
 \begin{align}
     \left[ \partial_u \bar{\phi}^{0a}(u,z,\bar{z}),\phi^{0b}(g,w,\bar{w})\right]=-i\gamma^{-1}_{z\bar{z}}\ell^{-2}\delta ^{2}( z-w) \delta ( u-g) \delta ^{ab}\, ,
 \end{align}
which leads to
\begin{align}
     \left[ Q^{-}_{H},\phi^{0d}_{k}(u,z_{k},\bar{z}_{k})\right]=Q_{k}g_{YM}C^{abd}\epsilon^a(z_{k},\bar{z}_{k})\phi^{0b}_{k}(u,z_{k},\bar{z}_{k})\, .
\end{align}
 One can find that the hard charge acts on in-state of the scalar particle linearly
 \begin{align}    Q_{H}^{-}\ket{\mathrm{in}}=\sum_{k=1}^{n}Q^{\mathrm{in}}_{k}T^a_k\epsilon^a(z_k^\mathrm{in},\bar{z}_k^\mathrm{in})\ket{\mathrm{in}}\,.
 \end{align}
Note that the commutation relation in \eqref{Acommutator} will yield the same result for the hard charge acting on in-state of the gluon. Finally, combining the actions of the soft and hard charges, the Ward identity in the coordinates space leads to a soft gluon theorem in dS spacetime
\begin{align}
     \lim_{\omega\rightarrow 0} \omega^2\mel**{\mathrm{out}}{a_{+}^{b}\mathcal{S}-\mathcal{S}a_{-}^{b\dagger}}{\mathrm{in}}&=i\frac{1+w\bar{w}}{2\sqrt{2}}\frac{R}{\ell^2}\left[\sum_{l=1}^{m}\frac{Q_l^{\mathrm{out}}T^b_l}{w-z_l^{\mathrm{out}}}-\sum_{k=1}^{n}\frac{Q_k^\mathrm{in}T^b_k}{w-z_k^{\mathrm{in}}}\right]\mel**{\mathrm{out}}{\mathcal{S}}{\mathrm{in}}\, .
\end{align}
Considering the null parametrization in \eqref{parametrization}, the soft theorem in momentum space is
\begin{align}
     \lim_{\omega\rightarrow 0} \mel**{\mathrm{out}}{a_{+}^{b}\mathcal{S}-\mathcal{S}a_{-}^{b\dagger}}{\mathrm{in}}&=\frac{i}{2\omega\ell}\left[\sum_{l=1}^{m}Q_l^{\mathrm{out}}\frac{T^b_l p_l^{\mathrm{out}}\cdot\epsilon}{p_l^{\mathrm{out}}\cdot p}-\sum_{k=1}^{n}Q_{k}^{\mathrm{in}}\frac{T^b_k p_k^{\mathrm{in}}\cdot\epsilon}{p_k^{\mathrm{in}}\cdot p}\right]\mel**{\mathrm{out}}{\mathcal{S}}{\mathrm{in}}\, .
\end{align}
It is amusing to see that the dS radius has exactly the same type of correction to soft photon and soft gluon theorem in dS spacetime. The flat limit of the soft gluon theorem is just the same as the soft photon theorem.


\section{Conclusion}

In this paper, a soft photon theorem and a soft gluon theorem are derived in dS spacetime from the Ward identity of the near horizon symmetry. Our results indicate that the cosmological constant affects the soft theorems universally. Both the soft photon and soft gluon theorem have a well defined flat limit. The soft theorems after the flat limit recover exactly the flat spacetime soft photon and soft gluon theorems.

As an ending remark, it is worth mentioning that we focus on massless hard particles in the soft theorem. While, the soft photon theorem derived in \cite{Bhatkar:2022qhz} involves only massive hard particles and the zero mass limit of the hard particles seems not well defined. So it is not clear to us how to connect our results in the present form to the ones in \cite{Bhatkar:2022qhz}.

\section*{Acknowledgments}
The authors thank Sayali Bhatkar, Diksha Jain, and Adam Tropper for useful correspondences.
This work is supported in part by the National Natural Science Foundation of China (NSFC) under Grants No. 11905156 and No. 11935009. K.Y.Z. is also supported in part by NSFC Grant No. 11905158.

\bibliography{ref}

\end{document}